\documentclass[10pt,floatfix,aps,prd,twocolumn,showkeys, preprintnumbers, nofootinbib, superscriptaddress,nolongbibliography]{revtex4-2}
\usepackage[utf8]{inputenc}
\usepackage[sort&compress]{natbib}
\usepackage{ulem}
\usepackage{overpic}
\usepackage{bm}
\usepackage{times}
\usepackage{amssymb,amsbsy,amsmath,amsfonts}
\usepackage{graphicx}
\usepackage{lipsum}
\usepackage{color}
\usepackage{booktabs}
\usepackage{makecell}
\usepackage{rotating}
\usepackage{srcltx}
\usepackage{slashed}
\usepackage{subfigure}
\usepackage{multirow}
\usepackage{verbatim}
\usepackage{hyperref}
\hypersetup{colorlinks,linkcolor=red,citecolor=red}
\usepackage{tabularx}
\usepackage{newtxtext, newtxmath}
\usepackage{physics}
\usepackage{tikz,import}
\usepackage{tikz-cd}
\usepackage{pgfplots}
\usepackage{CJKutf8}
\usepackage{bookmark}
\begin{document}
\title{Machine learning unveils the quark mass dependence of the pseudoscalar meson decay constants in three-flavour N$^2$LO ChPT}
\author{Zejian~Zhuang~(\begin{CJK*}{UTF8}{gbsn}庄泽坚\end{CJK*})}
\email{zejian.zhuang@uv.es}
\affiliation{Departamento de F\'{\i}sica Te\'orica and IFIC,
Centro Mixto Universidad de Valencia-CSIC, Parc Científic UV, C/ Catedrático José Beltrán, 2, 46980 Paterna, Spain} 
\author{Fernando~Gil-Dom\'{\i}nguez}
\email[Contributed equally to this work:~]{fgildominguez@uestc.edu.cn}
\affiliation{School of Physics, University of Electronic Science and Technology of China, Chengdu 611731, China}
\author{Raquel~Molina}
\email{raquel.molina@ific.uv.es}
\affiliation{Departamento de F\'{\i}sica Te\'orica and IFIC,
Centro Mixto Universidad de Valencia-CSIC, Parc Científic UV, C/ Catedrático José Beltrán, 2, 46980 Paterna, Spain}   
\begin{abstract}
The quark mass dependence of the pseudoscalar meson decay consntants, $f_\pi, f_K$ and $f_\eta$, are determined from three-flavor N$^2$LO ChPT till pion masses around $780$ MeV, near the SU(3) limit. This is done by conducting an analysis of recent LQCD data using the LASSO method, a machine-learning technique which allows to pin down the relevant low-energy-constants with high precision. Since the pion decay constant is a fundamental quantity which usually appears in relevant phenomenological lagrangians or Effective Field Theories based on QCD at low energies, this analysis can be used as input to evaluate the quark mass dependence of hadronic states. As an example, we predict the masses of the octect baryons in the SU(3) limit within covariant Baryon Chiral Perturbation Theory. 
\end{abstract}
\maketitle

%%
% \cm{RM. I have shortened the title. Let me know if it is okey with you.}
% \noindent{\it Introduction.---} 
\section{Introduction}
The pion decay constant is one of the most relevant physical magnitudes for processes in the low-energy regime in hadron physics. In Quantum Chromodynamics (QCD) it is defined in relation to the coupling of the pion to the axial current~\cite{ParticleDataGroup:2024cfk}
\begin{eqnarray}\label{eq:fpi}
    \bra{0}A_\mu(0)\ket{\pi^-(p)}=ip_\mu f_\pi; \quad A_\mu=\bar{u}\gamma_\mu\gamma_5 d\ .
\end{eqnarray}
It enters in the pion leptonic decay $\pi\to l\nu$, which at lowest order is\footnote{Up to electromagnetic corrections},
\begin{eqnarray}
    \Gamma^{(0)}=\frac{G_F^2 |V_{ud}|^2 f_\pi^2}{8\pi} m_\pi m_l^2\left(1-\frac{m_l^2}{m_\pi^2}\right)\ ,
\end{eqnarray}
% \lipsum[1-2]
where $m_\pi$ and $m_l$ are the pion and lepton mass, $V_{ud}$ the Cabibbo-Kobayashi-Maskawa (CKM) matrix and $G_F$ the Fermi coupling constant, and in the relation between the weak, $g_A$, and strong coupling constant, $g_{\pi N}$, through the Goldberger-Treiman relation~\cite{Goldberger:1958tr},
\begin{eqnarray}\label{eq:coupmp}
    f_\pi g_{\pi N}= \sqrt{2}m_p g_A\ ,
\end{eqnarray}
which are exact in the chiral limit. Moreover, the pion decay constant also enters in the scalar and vector pion form factor, in the weak decay of the $\eta$ to three pions, and in the leptonic kaon decay, so-called $K_{l_4}$~\cite{Gasser:1983yg,Bijnens:2009zd}. Furthermore, it dominates the coupling between pions and pions and nuclei in an Effective Field Theory (EFT) framework with chiral symmetry~\cite{Gasser:1983yg,Weinberg:1991um}. Experimentally, its current value is \cite{ParticleDataGroup:2024cfk},
\begin{eqnarray}\label{eq:fpiv}
    \sqrt{2}\,f_{\pi^+} = 130.2(1.2)\,\mathrm{MeV} ,
\end{eqnarray}
where most of the uncertainty comes from the $V_{ud}$ determination\footnote{Here and in the following we use the normalization of the decay constants with the $\sqrt{2}$ as in the above equation}. A  natural way to study hadron interactions at low energies is through EFT's. The most well-known example in the case of two pseudoscalar mesons interacting is Chiral Perturbation Theory (ChPT)~\cite{Weinberg:1978kz,Gasser:1983yg,Gasser:1984gg}, which is built upon the symmetries of the QCD Lagrangian, such as chiral symmetry, the underlying symmetry of QCD that takes place at vanishing quark masses, as an expansion in the mass and momenta of the pseudoscalar mesons, emerging when the chiral symmetry is spontaneously broken, as dictated by the Nambu-Goldstone's theorem~\cite{Nambu:1960xd,Nambu:1961tp,Nambu:1961fr,Goldstone:1961eq}.   In this
manner, the light-pseudoscalar observables are obtained as a
model independent expansion over the chiral scale $4\pi f_0 =1.2$ GeV, being $f_0$ the pion decay constant at the chiral limit.

At leading order (LO) the only parameter in ChPT is the coupling between the two pseudoscalar mesons, controlled by $f_\pi$. At higher orders, it depends on the values of the so-called low-energy-constants (LECs), which account for the underlying quark-gluon interactions that occur at higher energies and also renormalize the
loop diagrams. These can be fitted to reproduce the experiment~\cite{Gasser:1983yg,Gasser:1984gg}. At NLO ${\cal O}(p^4)$, three-flavor ChPT has about 10 LECs $L_i, i=1,10$, entering in the scattering of two pseudoscalar mesons and in their masses and decay constants, which can be determined through the experiment and LQCD~\cite{Gasser:1984gg}. However, at higher orders, while in two-flavour N$^2$LO there are four combinations of ${\cal O}(p^6)$ LECs which can be fixed through existing well-known observables related to the scalar and vector form factors~\cite{Bijnens:2014lea}, and the $\bar{l}_i, i=1,7$ LECs are also determined either through experiment or latticeQCD~\footnote{There are in total $7$ LECs of ${\cal O}(p^4)$ and $52$ of ${\cal O}(p^6)$ in the N$^2$LO two-flavour case.}
~\cite{Bijnens:2009pq}, the situation remains much more unclear in three-flavor N$^2$LO ChPT ${\cal O}(p^6)$~\cite{Fearing:1994ga,Bijnens:1999sh,Bijnens:1999hw,Bijnens:2006zp}. On the one hand, the number of LECs that appear in this case, $\sim90$~\cite{Bijnens:2009pq}, makes it unmanageable to achieve a precise determination of those, partly because of the high correlations between them. On the other hand, while it is clear that N$^2$LO corrections are necessary to reproduce, for example, LQCD data on the masses and pseudoscalar decay constants, the values of the NLO LECs change considerably in N$^2$LO fits, spoiling the convergence of the ChPT series. Large $N_c$ arguments suggest that some of the LECs, like $L_4$, $L_6$ and the combination $2L_1-L_2$, should be suppressed in this limit~\cite{Bijnens:2014lea}, what might not be always compatible with the best solution in N$^2$LO fits. Still, a reasonable expectation is that the NLO accounts for a correction of the order of $25\%$, while being $7\%$ for the N$^2$LO, and much smaller, $\sim 1.5\%$, for the N$^3$LO contribution~\cite{Bijnens:2014lea}. 

%Up to date, NLO ChPT is clearly insufficient to predict pseudoscalar-meson masses and decay constants for pion masses higher than $450$ MeV.
The first N$^2$LO ChPT calculation was performed in~\cite{Gasser:1983yg,Gasser:1984gg}, and there are available some reviews~\cite{Bernard:2006gx,Bijnens:2006zp,Bijnens:2009pq}. Moreover, there exists LQCD determinations employing two-flavor NLO ChPT to conduct chiral extrapolations of the pseudoscalar decay constants and other observables as the pion scalar radius, and the pion vector form factor~\cite{JLQCD:2008zxm,ETM:2009ztk,MILC:2010hzw,Borsanyi:2012zv,BMW:2013fzj,Boyle:2015exm,Brandt:2013dua,Gulpers:2015bba}. There are available few two-flavour N$^2$LO fits~\cite{JLQCD:2008zxm,Mawhinney:2009jy,Scholz:2011rk,Brandt:2013dua}, and also three-flavours NLO ChPT fits~\cite{Beane:2006kx, MILC:2009ltw,PACS-CS:2008bkb,RBC-UKQCD:2008mhs,Bijnens:2011tb}. See also~\cite{FlavourLatticeAveragingGroupFLAG:2021npn,FlavourLatticeAveragingGroupFLAG:2024oxs}.  Most of these simulations are performed at unphysical pion masses lying in a chiral trajectory set up at the physical strange quark mass, reaching the three-flavour symmetric line $m_q=m_{s,\mathrm{phys}}$, $q=u,d$ at a pion mass $m_\pi\sim 700-800$ MeV. 
\begin{figure}[!ht]
	\centering
	\includegraphics[width=0.95\columnwidth]{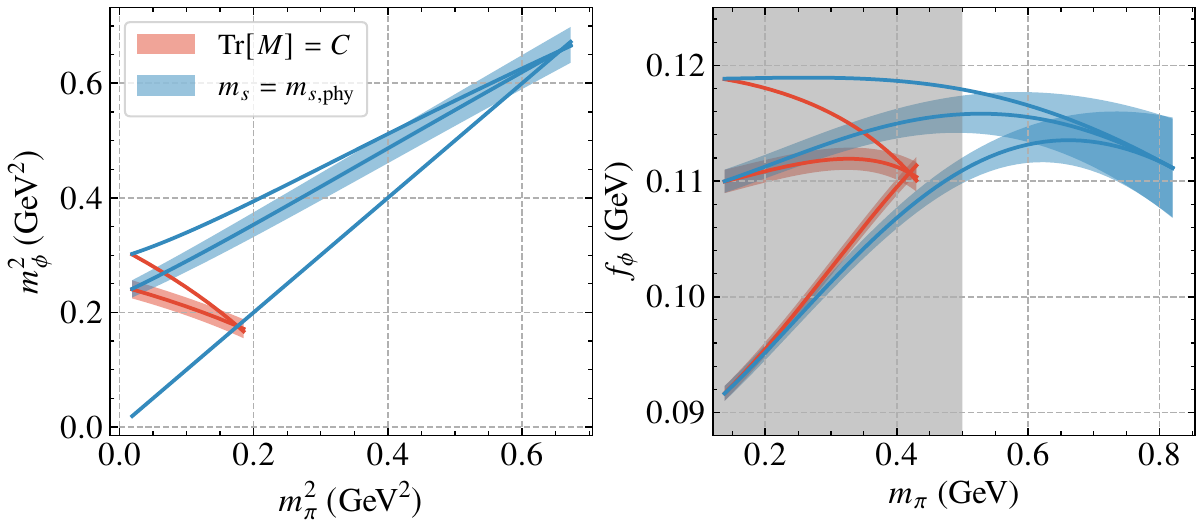}
	\caption{\label{fig:chpt-nlo}The pseudoscalar decay constants in NLO ChPT, that is frequently used till $m_\pi\simeq450$~MeV. The parameters $L_{ir}$, $i=1,\,8$ are taken from Ref.~\cite{Molina:2020qpw}.}
\end{figure}

One of the most difficult problems in non-perturbative QCD is that of reproducing low-energy excitations of the hadronic states. The combination of EFT's with unitarity techniques allows extending the range of applicability of the EFT. A popular example is the chiral unitary approach~\cite{Truong:1988zp,Dobado:1989qm,Oller:1998hw,Nieves:1998hp,GomezNicola:2001as}, that has been applied to reproduce resonances like the $\sigma$, $\kappa$ or $\rho$ meson in the meson sector, and the $N^*(1535)$, $\Lambda(1405)$, in the baryon sector~\cite{Kaiser:1995eg,Kaiser:1996js,Oset:1997it,Nieves:2001wt,Jido:2003cb}. Furthermore, the predictions of the chiral unitary approach have been tested by LQCD simulations, or applied in combination to study hadron-hadron scattering at low energies, successfully reproducing some of the resonances appearing in the light-meson spectrum~\cite{Hanhart:2008mx,Molina:2015uqp,Guo:2016zos,Hu:2016shf,Guo:2018zss,Nebreda:2010wv,Mai:2019pqr,Molina:2020qpw,Fischer:2020yvw,Niehus:2020gmf,Pelaez:2021dak,Zhuang:2024udv}. In this regard, a relevant question is the connection of these predictions with QCD at the symmetric point, where the SU(3) symmetry holds. This can be especially relevant in the case of two-pole structures~\cite{Jido:2003cb,Meissner:2020khl,Zhuang:2026lta}. The predictions from the chiral unitary approach can be tested by LQCD by performing simulations at different quark masses, including the three-flavour symmetric point. Up to date, the chiral unitary approach faces difficulties in this respect, since, assuming one utilizes the most frequently used three-flavor NLO ChPT, this cannot be applied at pion masses higher than $450$ MeV, see Fig.~\ref{fig:chpt-nlo}. Thus, it is desirable to extend the range of applicability of three-flavor ChPT by including higher order terms, testing also the convergence of the ChPT series around the symmetric point. 

Thus, in this work, we analyze the available LQCD data at present on pseudo-scalar meson masses and decay constants with three-flavour N$^2$LO ChPT. As a novelty, and to handle the large number of LECs that appear at this order, we employ, for the first time in this context, a machine learning algorithm, the Least Absolute Shrinkage and Selection Operator (LASSO) method~\cite{orilasso,Lasso,isl}, which is useful to disentangle the correlations and that provides a predictive solution by disregarding unnecessary free parameters. 

This article continues as follows. In the next section, we provide the main details for the N$^2$LO calculation and the fit performed with the LASSO method. Then, we provide our results and conclusions.

\section{Formalism}
% \noindent{\it Formalism. ---}
The masses and decay constants of the pseudoscalar mesons are evaluated within ChPT at ${\cal O}(p^6)$ in \cite{Amoros:1999dp}. The masses of the pseudoscalar mesons can be evaluated from the position of the pole in a two-point green function that contains the relevant particle as the intermediate state. The full formula for the squared mass of the pseudoscalar meson $a=\pi,K,\eta$ can be expressed at ${\cal O}(p^6)$ as,
\begin{eqnarray}\label{eq:mass}
    m^2_a=m^2_{0a}+m^{2\,(4)}_a+m^{2\,(6)}_{ct,a}+m^{2\,(6)}_{loop,a}\ ,
\end{eqnarray}
where the first term of the right-hand side stands for the bare mass, the second is the full ${\cal O}(p^4)$ contribution, which depends on the $L_i's$, concretely, the combinations, $2L_6-L_4$, $2L_8-L_5$, for the pion and kaon mass, and $L_i, i=4,8$ for the $\eta$ mass. The last two terms in Eq.~(\ref{eq:mass}) are the polynomial or counter term ($ct$) and loop ${\cal O}(p^6)$ contributions, respectively. The $ct$ term is linear on the ${\cal O}(p^6)$ LECs, the $C_i$'s, $i=12-21,31-33$, where $C_{18}$ and $C_{33}$ only appear in the eta mass. The  ${\cal O}(p^6)$ loop part is built upon ${\cal O}(p^2)$, ${\cal O}(p^4)$ vertices, and therefore, it depends on $L_i$, $i=1,8$ and products of the type $L_iL_j$. The regularization of the loops in \cite{Amoros:1999dp} is done in the $\overline{MS}$ scheme. The specific NNLO formulas for the different terms appearing in Eq.~(\ref{eq:mass}) are given in the appendix A1 of~\cite{Amoros:1999dp}. The decay constant of a pseudoscalar meson is defined as in Eq.~(\ref{eq:fpi}) but replacing $\pi^-$ by the $\phi^a$, the axial current and $f_\pi$ by $A_\mu^a$ and  $f_a$ of the corresponding pseudoscalar meson. The full ${\cal O}(p^6)$ formula reads as,
\begin{eqnarray}\label{eq:dec}
    f_a=f_0(1+\bar{f}_a^{(4)}+\bar{f}_{ct,a}^{(6)}+\bar{f}_{loop,a}^{(6)})\ ,
\end{eqnarray}
where $f_0$ is the pseudoscalar decay constant in the chiral limit, $\bar{f}_a^{(4)}$ contains the full ${\cal O}(p^4)$ contribution, depending on $L_4$ and $L_5$, and $\bar{f}_{ct,a}^{(6)}$, $\bar{f}_{loop,a}^{(6)}$ stand for the ${\cal O}(p^6)$ $ct$ and loop contribution, respectively. The $\bar{f}_{ct,a}^{(6)}$ part is linear in the ${\cal O}(p^6)$ LECs, $C_i$, $i=14-18$, where $C_{18}$ appears only in the eta decay constant.  Similarly to the mass case, $\bar{f}_{loop,a}^{(6)}$ depends linearly on $L_i$ and on $L_iL_j$ products. The explicit formulas for the various contributions in Eq.~(\ref{eq:dec}) can be found in the appendix $A2$ of~\cite{Amoros:1999dp}. The N$^2$LO formulas for the pseudoscalar meson masses and decay constants depend, in total, on two parameters for the LO, $m_0^2, f_0$, $8$ NLO LECs ($L_i$), and $13$ N$^2$LO LECs ($C_i$). Thus, there are $23$ free parameters in the N$^2$LO formulas. Here, we fix $f_0=80$~MeV. 

The LASSO method is a procedure for selecting the optimal solution in a class of linear models~\cite{orilasso,Lasso,isl}. By optimal we refer to a solution involving a subset of parameters that describes the data with enough accuracy, neglecting those that do not reduce significantly the $\chi^2$. The number of parameters is regulated by introducing a new parameter, $\lambda$, in the $\chi^2$, as follows,
\begin{eqnarray}\label{eq:chi2}
\chi^2(\vec{L},\vec{C})=\sum_i\frac{(y_i-f_i(\vec{L},\vec{C}))^2}{\sigma^2_i}+\frac{\lambda}{10}\sum_j|C_j|\ .
\end{eqnarray}
Here, $\vec{L}$ and $\vec{C}$ are the vectors containing the NLO and N$^2$LO LECs, respectively. The $y_i$ and $f_i$ functions are the data and fitting functions for the pseudoscalar mass and decay constants. The penalty term is introduced only for the N$^2$LO LECs included in the fit. In order to choose the optimal value of $\lambda$, the data are divided into three subsets, two of them chosen for the training set, $\chi^2_T$, and the third one for the validation set, $\chi^2_V$. While $\chi^2_T$ increases monotonically with $\lambda$, $\chi^2_V$ is expected to have a sweet spot where the data are not underfitted, not overfitted. This $\lambda$ selection procedure is called \textit{cross validation}. In addition, information criteria, AIC, AICc and BIC, can be used to select the optimal $\lambda$~\cite{Gil-Dominguez:2023eld}. Any of these criteria penalizes models with a large number of parameters. 

In this work, we take as input the LQCD data for pseudoscalar meson masses and decay constants from MILC~\cite{MILC:2009mpl}, UKQCD~\cite{RBC:2014ntl} and CLS~\cite{Bruno:2016plf,Ce:2022kxy,Bali:2022qja} ensembles, which lie on the chiral trajectories $\mathrm{Tr}[M]=C$, $m_s=m_{s,\mathrm{phy}}$, and $m_s=m_{u,d}$. See also~\cite{RQCD:2022xux} for the scale setting of the data in~\cite{Bali:2022qja}. \\
\begin{table*}[!ht]
    \renewcommand{\arraystretch}{2.2}
     \setlength{\tabcolsep}{0.37cm}
    \centering
    \caption{\label{tab:hadron-mass}The masses of the psudo-scalar mesons and the decay constants up to NNLO at physical pion mass and the SU3 flavor. The LECs are fixed by the Fit 2.A. The unit in MeV. On the first row, we take isospin average mass of kaon and the decay constants are re-evaluated in our renormalization.}
    \begin{tabular}{lccccccc}
    \hline\hline
    ~ & $m_K$ & $f_\pi$ & $f_K$ & $f_K/f_\pi$ &
    $m_\phi^{(\mathrm{SU3})}$ & $f_\phi^{(\mathrm{SU3})}$ \\ \hline
    PDG~\cite{ParticleDataGroup:2024cfk} & $496$ & $92.07(85)$ & $110(2)$ & $1.193(2)$ & $\cdots$ & $\cdots$ \\
    \(m_s=m_{s,\mathrm{phy}}\)~(NLO)~\cite{Molina:2020qpw} & \(491(15)\) & \(91.6(7)\) & \(110(1)\) & \(1.201(12)\) & \(800(19)\) & \(112(4)\)\\
    \(m_s=m_{s,\mathrm{phy}}\)~(NNLO) & \(496(2)(1)\) & \(92.3(11)(6)\) & \(110.2(9)(12)\) & \(1.193(8)(16)\) & \(780(60)(22)\) & \(128(4)(10)\)\\
    \(\mathrm{Tr}[M]=C\)~(NNLO) & \(494(2)(1)\) & \(92.3(10)(7)\) & \(110.1(13)(7)\) & \(1.193(17)(5)\) & \(420(3)(1)\) & \(110(1)(12)\)
    \\\hline\hline
\end{tabular}
\end{table*}

\section{Results}
% \noindent{\it Results.---}%
We conduct two different fits of the pseudoscalar meson masses and decay constants with ChPT as described below:
\begin{itemize}
    \item Fit 1: A global fit of data with the NLO ($L_{ir}$'s) and N$^2$LO ($C_{ir}$'s) LECs as free parameters,
\end{itemize}
or the LASSO method is applied to the N$^2$LO LECs ($C_{ir}$'s), following these two different strategies:
\begin{itemize}
        \item Fit 2A: We first set $C_{ir}=0$, and fit the pseudoscalar meson masses and decay consants to obtain the $L_{ir}$'s with the NNLO formulas including up to two loops. Then, we fix the $L_{ir}$ and conduct a refit to get the $C_{ir}$'s.
        \item Fit 2B: We leave the $C_{ir}$'s as free parameters $C_{ir}\neq 0$ and apply the LASSO method directly. 
\end{itemize}

Some of the LECs in this analysis were fixed to reproduce some known observables. Concretely, $L_{1r}$ and $L_{2r}$ are restricted to, $2L_{1r}-L_{2r}=0.36\times 10^{-3}$, in order to reproduce the $\rho$ meson mass in $\pi\pi$ scattering within unitarized chiral perturbation theory for pion masses in the range of $m_\pi\sim200-450$ MeV in LQCD simulations~\cite{Molina:2020qpw}. The parameters which determine the chiral trajectories, $\mathrm{Tr}[M]=C$ and $m_s=m_{s,\mathrm{phy}}$, are adjusted to the LQCD data as done in~\cite{Zhuang:2024udv}. These values correspond to $B_0 C=488^2\mathrm{MeV}^2$ and $B_0  m_{s,\mathrm{phy}}=474^2\mathrm{MeV}^2$, with $B_0$ the constant related to the quark condensate at the chiral limit, $\Sigma_0=-\bra{0}q\bar{q}\ket{0}_0=B_0f_0^2$. Finally, since the $C_{18r}$ and $C_{33r}$ enter only in the $\eta$ mass and decay constant, we have tuned these parameters after conducting the fits to obtain the physical values of $m_\eta$ and $f_\eta$.
\begin{figure}[!htbp]
	\centering
	\includegraphics[width=\columnwidth]{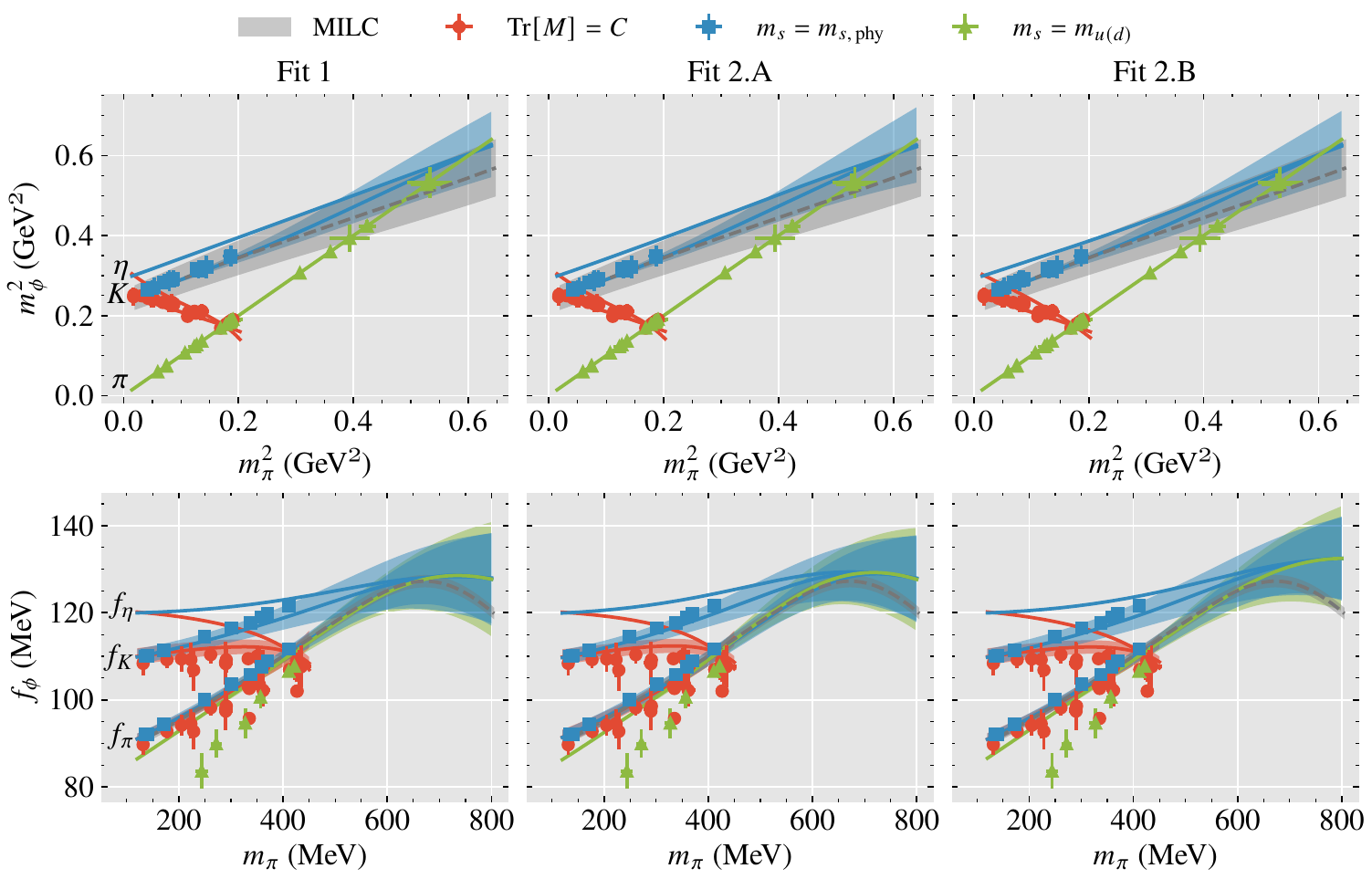}
	\caption{The LQCD data with trajectories $\mathrm{Tr}[M]=C$ and $m_s=m_{s,\mathrm{phy}}$~\cite{Ce:2022kxy,RQCD:2022xux,RBC:2014ntl,Bruno:2016plf}, and SU3~\cite{Bali:2022qja} used as inputs to constrain the LECs up to NNLO. The circles, squares and the triangles denote the $\mathrm{Tr}[M]=C$, $m_s=m_{s,\mathrm{phy}}$, and the $m_s=m_{u,d}$, respectively. The reduced-$\chi^2$ for every fit is about 1.}
    \label{fig:dec}
\end{figure}
 \begin{figure}[!ht]
 	\centering
 	\includegraphics[width=\columnwidth]{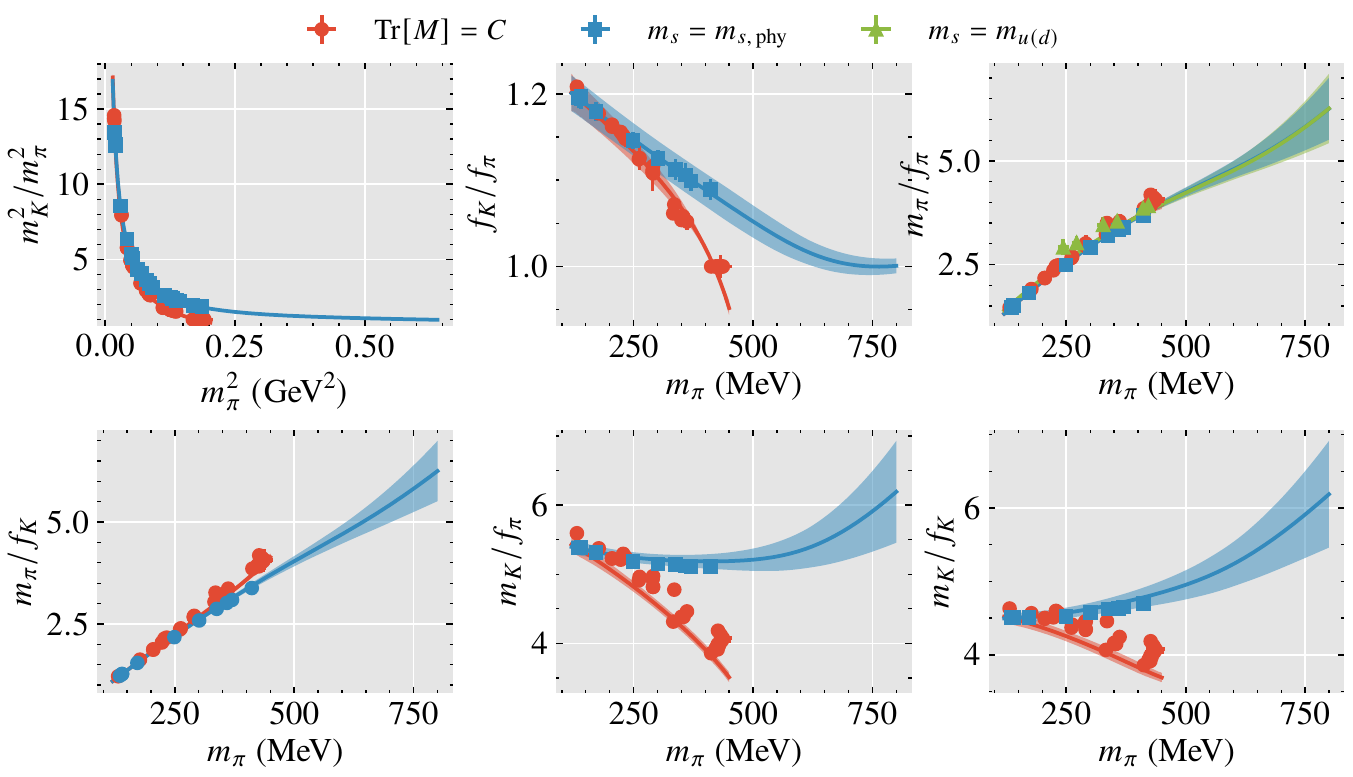}
 	\caption{\label{fig:ratio}The ratios $m_K^2/m_\pi^2$, $f_K/f_\pi$, $m_\pi/f_\pi$, $m_\pi/f_K$, $m_K/f_\pi$, and $m_K/f_K$ obtained from the Fit~2.A.}
 \end{figure}

\begin{figure}[!ht]
    \centering
    \includegraphics[width=0.9\columnwidth]{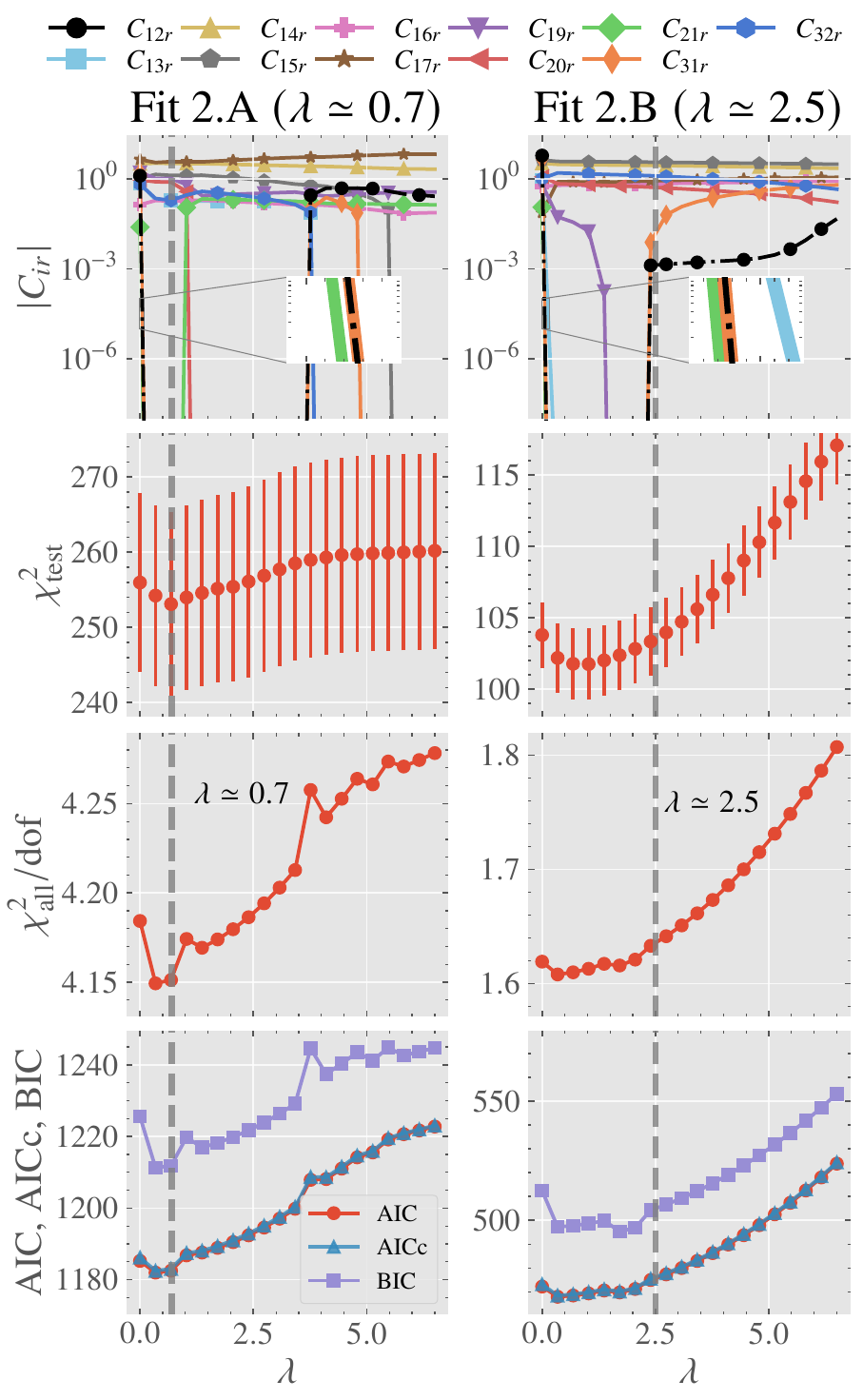}
    \caption{Results from the LASSO regression method. Left plots correspond to the fit with $C_{ir}=0$, and right ones to taking $C_{ir}\neq 0$. The dashed gray line shows the optimal value for the $\lambda$ parameter that obtained in each of the cases.}\label{fig:cvaic}
\end{figure}

\begin{figure}[!htb]
	\centering
	\includegraphics[width=0.9\columnwidth]{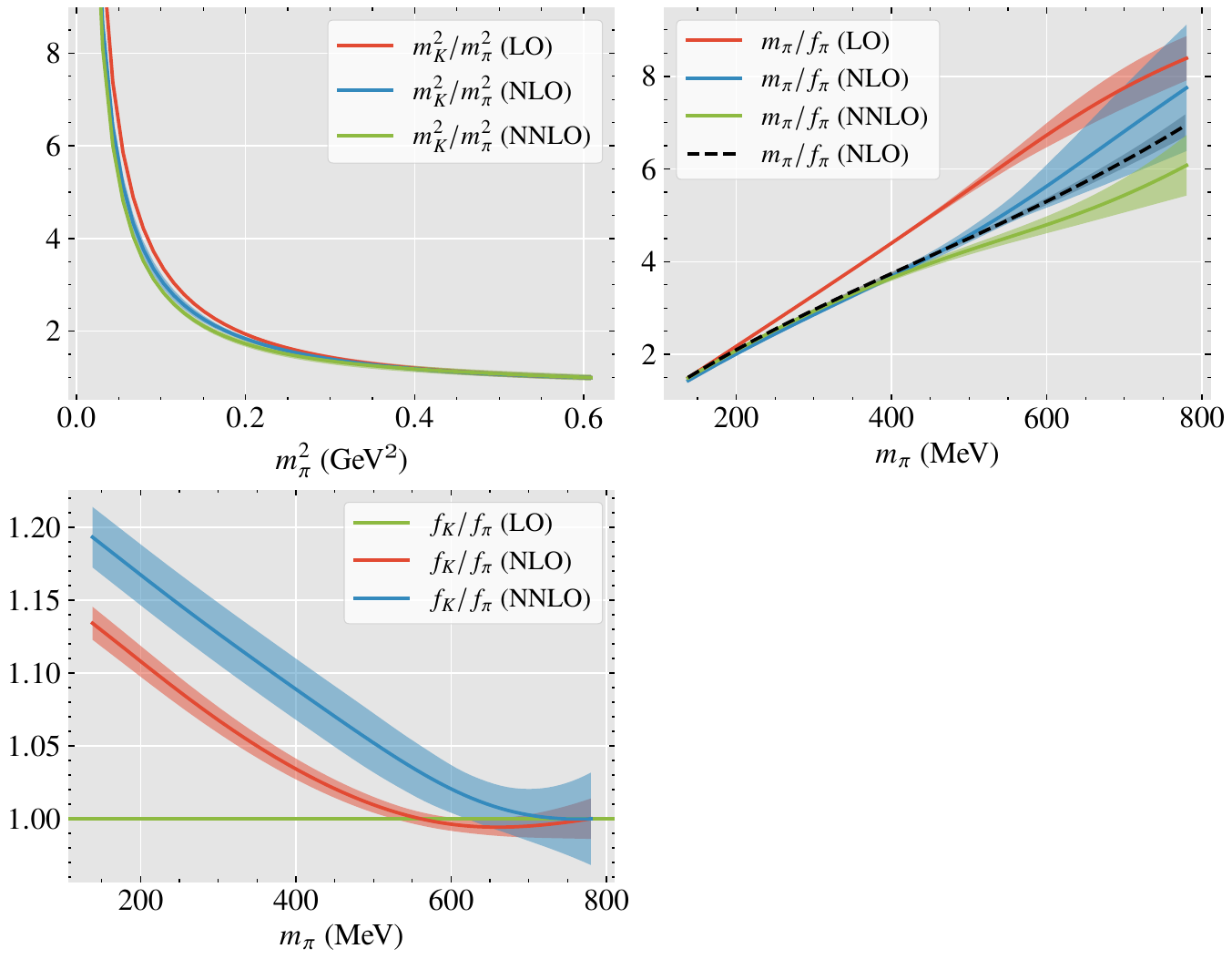}
    % \resizebox{0.9\columnwidth}{!}{\import{}{figure/convergence.pgf}}
	\caption{\label{fig:convengence}The ratios $m_K^2/m_\pi^2$, $f_K/f_\pi$, and $m_\pi/f_\pi$ at LO, NLO and NNLO obtained from the Fit~2A. The ratio $m_\pi/f_\pi$ at NLO is taken from Ref.~\cite{Molina:2020qpw}.}
\end{figure}

The pseudoscalar meson masses and decay constants obtained in the various fits, as well as the pertinent ratios, are shown in Figs.~\ref{fig:dec} and~\ref{fig:ratio}, respectively. As can be seen, the N$^2$LO fits conducted provide very similar results for these observables, emphasizing the stability of our results. In Fig.~\ref{fig:convengence} we also show the comparison between the performance of NLO and N$^2$LO fits for the ratios of squared masses and decay constants. From this figure one sees that the relative difference in the ratio $f_K/f_\pi$ between NLO and N$^2$LO is only of $5$~\%. %Note that the interaction in the chiral lagrangian for two pseudoscalar mesons is dimensionless, and therefore, only ratios of masses and decay constants do enter in the evaluation.
In this work, at N$^2$LO, the SU(3) flavor symmetric point is reached at,
\begin{equation}
    m_\pi=780(60)(22)\,(\mathrm{MeV}).
\end{equation}
From Fig.~\ref{fig:dec}, we can see that the chiral trajectories obtained now for pion masses above $450$~MeV with N$^2$LO ChPT works well till reaching the SU(3) limit. Hence, the analysis presented here for the pseudoscalar meson decay constants can be used as input for the calculation of any observable depending of those at these high pion masses. In Table~\ref{tab:hadron-mass}, we provide the result for $m_P, f_P$, $P=\pi, K, \eta$ at NLO and N$^2$LO compared to the experimental values. We observe that the N$^2$LO fit, although it has smaller impact for lower pion masses, it improves the precision of the masses and decay constant data compared to experiment.\\
The results for the N$^2$LO LECs obtained in the different fits are shown in Tables~\ref{tab:lir} and \ref{tab:cir}. Remarkably, the ML algorithm LASSO unveils that some of the LECs can be set to zero, producing optimal fits as if these are let to be free. Depending on how the fit is performed (fixing $L_{ir}$ from a fit with $C_{ir}=0$ and/or applying the LASSO), different  N$^2$LO parameters disappear, $C_{12r}, C_{21r}$ and $C_{31r}$, in Fit 2A, or $C_{13r}, C_{19r}$ and $C_{21r}$ in Fit 2B. This is due to the correlations between the LECs. Note that two of the LECs that vanish in Fit 2A, $C_{12r}$ y $C_{31r}$, also vanish in Fit 2B for $\lambda\sim 1$, however, these come back for higher values of $\lambda\sim2.5$, and we take here the conservative option by keeping those. But the main achievement of the ML algorithm is that, the number of NNLO LECS necessary to describe the present LQCD data is $11-3=8$\footnote{We do not count here $C_{18r}$ and $C_{33r}$ since these enter only in the $\eta$ mass and decay constant, which have been fixed.}. This decision is based on the combination of the LASSO method with cross-validation or information criteria. See Fig.~\ref{fig:cvaic}. In the first row of this figure, it can be seen that some of the LECs vanish for certain values of the learning parameter, $\lambda$. In the second to fourth rows, $\chi^2_\mathrm{test}$, $\chi^2_{\mathrm{all}}/\mathrm{dof}$ and the information criteria, $AIC, AICc, BIC$, are depicted, respectively. The $\chi_\mathrm{test}^2$ and the criteria show a clear minimum around $\lambda\sim 1$. Here we opt by choosing the optimal $\lambda$ as the average $\lambda$ from the cross validation and information criteria decision, and such that $R(\lambda_{\mathrm{opt}})=R(\lambda_{\mathrm{min}})+\Delta$, with $R=\chi^2_\mathrm{test}, AIC, BIC, BICc$ and $\Delta$ the uncertainty of $R$ at $\lambda_{\mathrm{min}}$ ($1-\sigma$ rule) and taking the test set as the validation set. This corresponds to $\lambda\simeq 0.7$ for Fit 2A, and $\lambda\simeq 2.5$ for Fit 2B. On the one hand, the numerical values of the NLO LECs obtained in the different N$^2$LO fits, Table~\ref{tab:lir}, are similar, and even compatible for some of the LECs, as $L_{4r}, L_{5r}$ and $L_{7r}$ in Fits 2A and 2B, or $L_{7r}$ and $L_{8r}$ in Fits 1 and 2B. On the other hand, we see that, while $C_{14r}, C_{15r}, C_{16r}$ take similar values in the three fits, the numerical values of the rest of the NNLO LECs, see Table~\ref{tab:cir}, can vary significantly from one fit to another, even though the results for the fits concerning the masses and decay constants are quite stable, indicating that, taking into account the correlations between these parameters is crucial. The correlation matrices between the NLO and NNLO ChPT LECs are provided in this article for the first time; see Appendix~\ref{app:cor}.

Note also that even though the N$^2$LO ChPT series has a large number of LECs at this order ($1+6+13=20$),  some of the LQCD data cannot be fitted well, regardless of the three vanishing parameters in the LASSO method are considered or not. See~Figs.~\ref{fig:dec} and~\ref{fig:ratio}. While the LQCD data in the $m_s=m_{s,\mathrm{phys}}$ trajectory are perfectly reproduced, some of the data of the masses and decay constants over the $\mathrm{Tr}M=C$ and $m_s=m_{u,d}$ curves, are not well covered by the fit results. However, note that these clouds of data points show clearly some inherent tensions between the statistical uncertainties, reflecting the lack of the consideration of possible systematics effects. These tensions in the data and with the fit results, are present even for not very high pion masses around $200-400$ MeV, where even the NLO fits are supposed to perform well, revealing some possible issues with LQCD simulation artifacts. Still, the description of the pseudoscalar meson masses, decay constants and ratios, is quite optimal overall.

\begin{figure}[!hbtp]
	\centering
	\includegraphics[width=0.9\columnwidth]{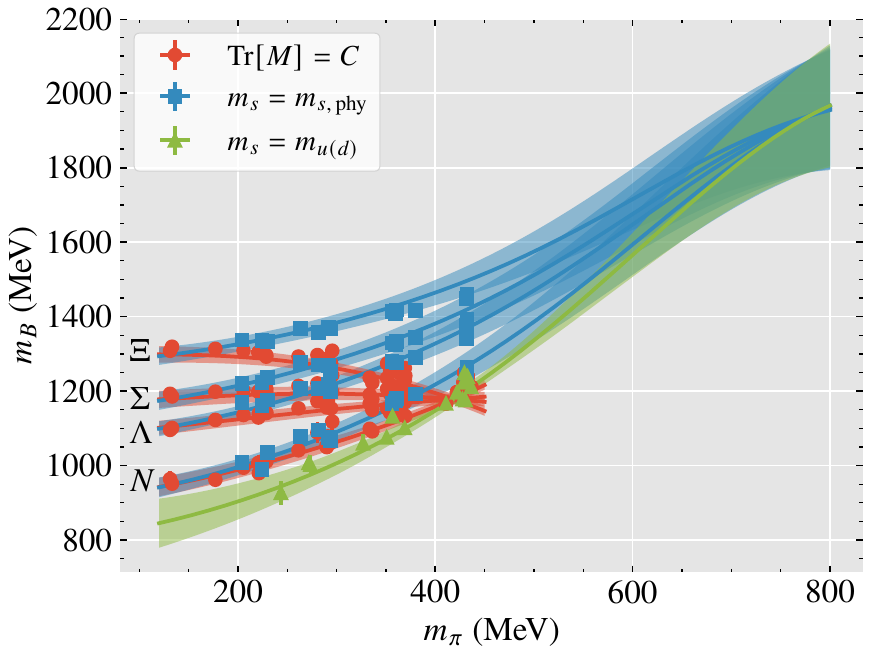}
	\caption{\label{fig:baryon}Pion mass dependence of the octet baryon masses predicted in this work beyond $m_\pi=450$~MeV by taking as input the result of the analyses of the pseudoscalar meson masses and decay constants at N$^2$LO, and the LQCD data for the baryon masses from the RQCD Collaboration~\cite{RQCD:2022xux}. See also Ref.~\cite{Zhuang:2024udv}. The SU(3) symmetry point of the baryon octet mass over \(\mathrm{Tr}[M]=C\), and \(m_s=m_{s,\mathrm{phy}}\) are \(1180(12)\)~MeV, and \(1938(160)\)~MeV, respectively.}
\end{figure}

Finally, in Fig.~\ref{fig:baryon} we show our predictions for the baryon masses calculated using one-loop covariant Baryon Chiral Perturbation Theory~\cite{MartinCamalich:2010fp} using as input the pseudoscalar meson masses and decay constants evaluated here at N$^2$LO by reanalizing the LQCD data for the baryon masses in~\cite{RQCD:2022xux}. This analysis was previously done with the NLO input for the mesons in~\cite{Zhuang:2024udv} and the baryon masses obtained in the various chiral trajectories could only be employed till a pion mass of around $450$~MeV. With the present N$^2$LO ChPT results as input for the baryon masses, one can see that their trends when increasing the pion mass approaching the symmetric SU(3) point ($m_\pi\sim 780$~MeV) are quite reasonable and also describe well the lattice data. 
\begin{table*}[!ht]
\renewcommand{\arraystretch}{2.2}
    \setlength{\tabcolsep}{0.4cm}
    \centering
 \caption{\label{tab:lir}The LECs $L_{ir}\times10^{3}$\footnote{We follow the convention of Ref.~\cite{Molina:2020qpw}.}  obtained from the different fits. The first uncertainty is statistical, and the second one is systematic from the lattice spacing~\cite{RQCD:2022xux}.}
 \begin{tabular}{lcccccc}
\hline\hline
~ & $L_{2r}$ & $L_{4r}$ & $L_{5r}$ & $L_{6r}$ & $L_{7r}$ & $L_{8r}$\\
\hline
~\cite{Molina:2020qpw} & \(\cdots\) & \(-0.08(3)\) & \(0.98(7)\) & \(0.24(8)\) & \(0.008(140)\) & \(0.098(110)\)\\
Fit 1 & $0.60(3)(1)$ & $-0.04(2)(1)$ & $0.653(56)(30)$ & $0.192(13)(1)$ & $-0.20(17)(4)$ & $-0.05(3)(2)$\\
Fit 2.A & $0.498(37)(9)$ & $-0.075(20)(11)$ & $0.42(5)(5)$ & $0.134(12)(13)$ & $-0.25(17)(5)$ & $0.01(3)(2)$\\
Fit 2.B & $0.697(37)(11)$ & $0.083(20)(11)$ & $0.55(5)(5)$ & $0.26(13)(1)$ & $0.02(16)(5)$ & $-0.174(29)(22)$\\
%Fit 2.B2 & $0.653(2)(4)$ & $0.104(2)(3)$ & $0.222(4)(9)$ & $0.343(1)(2)$ & $0.283(11)(25)$ & $-0.424(2)(6)$\\
\hline\hline
\end{tabular}
\end{table*}
\begin{table*}
  \centering
\caption{\label{tab:cir}The values of the N$^2$LO parameters, $f_0^2C_{ir}\times 10^{6}$ ~(see also Table~\ref{tab:lir}). The LECs $C_{18r}$ and $C_{33r}$ are fixed to reproduce $m_\eta$ and $f_\eta$ at the physical pion mass.}
\renewcommand{\arraystretch}{2.2}
\setlength{\tabcolsep}{0.4cm}
\begin{tabular}{ccccccc}
\hline\hline
$C_{12r}$ & $C_{13r}$ & $C_{14r}$ & $C_{15r}$ & $C_{16r}$ & $C_{17r}$ & $C_{18r}$ \\\hline
$2.89(84)(22)$ & $-1.67(39)(11)$ & $-6.68(85)(24)$ & $2.94(74)(23)$ & $1.59(32)(9)$ & $1.31(20)(90)$ & $-1.3(14)$ \\
$\cdots$ & $-0.73(42)(12)$ & $-4.91(90)(30)$ & $2.0(8)(3)$ & $1.43(34)(12)$ & $4.1(21)(9)$ & $-4.0(2)$ \\
$2.2(9)(2)$ & $\cdots$ & $-5.6(9)(3)$ & $3.86(8)(26)$ & $1.0(3)(1)$ & $0.1(20)(9)$ & $-0.4(14)$ \\
\hline
$C_{19r}$ & $C_{20r}$ & $C_{21r}$ & $C_{31r}$ & $C_{32r}$ & $C_{33r}$ & \\
\hline
$0.15(31)(8)$ & $-0.12(20)(5)$ & $0.07(9)(2)$ & $-1.87(84)(22)$ & $0.33(39)(11)$ & $2.3(5)$ & \\
$-1.3(3)(1)$ & $0.65(22)(6)$ & $\cdots$ & $\cdots$ & $-0.32(42)(11)$ & $1.8(5)$ & \\
$\cdots$ & $0.05(22)(6)$ & $\cdots$ & $-3.0(9)(2)$ & $2.4(4)(1)$ & $-0.6(5)$ & \\
\hline\hline
\end{tabular}
\end{table*}

% \begin{table}
% 	\centering
% 	\renewcommand{\arraystretch}{1.5}
%      \setlength{\tabcolsep}{0.9cm}
% 	\caption{\label{tab:Cr1833}The LECs $C_{18r}$ and $C_{33r}$ in unit of $f_0^2C_{ir}\times 10^{6}$ used to fix the $m_\eta$ and $f_\eta$ at the physical pion mass in different fitting strategies.}
% 	\begin{tabular}{ccc}
% 		\hline\hline
% 		~ & $C_{18r}$ & $C_{33r}$\\\hline
% Fit 1 & $-1.3(14)$ & $2.3(5)$\\
% Fit 2.A & $-4.0(2)$ & $1.8(5)$\\
% Fit 2.B & $-0.4(14)$ & $-0.6(5)$\\
% %fit 2.B2 & $-0.53(3)$ & $-2.13(1)$\\
%         \hline\hline
% 	\end{tabular}
% \end{table}

% \noindent{\it Conclusions.---}%
\section{Conclusions}
For the first time, the pseudoscalar meson masses and decay constants have been determined till pion masses around the SU(3) symmetric point, reached here around $m_\pi=780$ MeV. This is done by conducting a fit of LQCD data in different chiral trajectories, $m_s=m_{s,\mathrm{phys}}$, $\mathrm{Tr}M=C$, and $m_s=m_{u,d}$, within three-flavor N$^2$LO ChPT applied in combination with ML methods. Concretely, the LASSO algorithm is able to reduce the number of parameters at N$^2$LO without losing quality in the observables investigated. This tells us that with the present data, three of the $13$ LECs at this order can be disregarded when reproducing the pseudoscalar meson masses and decay constants. While in analyses with some chiral-based EFT theories, certain arguments or hypothesis are often assumed to get rid of some of those because of the large number of free parameters, like the resonance saturation hypothesis~\cite{Bijnens:2014lea}, the ML algorithm used here is able to reduce the number of degrees of freedom without taking any assumption. This comes out naturally from the data used, minimizing the complexity of the general formalism according to the data needs. %The way the fit is performed can influence the numerical results for the LECs leading to numerical values which are different at the same order\footnote{For example, first adjusting the $L_{ir}'s$ with $C_{ir}'s=0$, and then fitting the NNLO LECs, rather than leaving all as free parameters.}, especially in the case of three-flavor NNLO ChPT, characterized by a large number of free parameters. 
We have also shown that once the correlations between the remaining parameters are properly accounted for, the pseudoscalar meson masses and decay constants are compatible in the different fit solutions provided here, and their precision can also be determined, reinforcing the stability of our results. The correlation matrices between the NLO and N$^2$LO ChPT parameters are yielded here for the first time. The quark mass dependence of the observables investigated can be used as input in any chiral based EFT theory that sets the scale through the pseudoscalar meson decay constants, like $f_\pi$, when observables depending on these quantities need to be determined for high pion masses, as it happens close to the SU(3) flavor symmetry point. In this regard, we have evaluated the octect baryon masses in various chiral trajectories approaching this limit within covariant Baryon Chiral Perturbation Theory. In the future, this input can also be used to extrapolate scattering data from LQCD simulations using EFT methods. As for example, it has been done recently for $D\pi$ scattering,~\cite{Zhuang:2026lta}, being a useful tool to extract properties of resonances from LQCD and unveil the nature of two-pole structures near the SU(3) limit.

% \noindent{\it Acknowledgments.---}%
% \section*{Acknowledgments}
\begin{acknowledgments}
We are grateful to J. Bijnens for the three-flavor N$^2$LO ChPT formulas and acknowedge to  the MILC and RQCD Collaboration for making the data available to us. Z.~Zhuang is grateful to J.~R.~Pel\'aez and Yao~Ma for useful discussions. Z.~Zhuang also thanks Kostas~P. for his kind encouragement and warm support. The plots were made with \texttt{Matplotlib}~\cite{Hunter:2007} and the calculations were performed by \texttt{Julia}~\cite{Julia-2017}. R.~M. acknowledges support from the ESGENT program with Ref. ESGENT/018/2024 and the PROMETEU program with Ref. CIPROM/2023/59, of the Generalitat Valenciana.  This work is supported by the Ministerio de Ciencia e Innovación, research contract PID2023-147458NB-C21,
funded by MICIU/AEI/10.13039/501100011033. This project has also received funding from the European Union Horizon 2020 research
and innovation program under the program H2020-INFRAIA-2018-1, grant agreement No. 824093 of the STRONG-2020.
\end{acknowledgments}

% \clearpage
\appendix
\section{\label{app:cor}Correlations}
We provide the correlation matrices in the three fits.
\begin{figure*}[!hpt]
    \centering
    \includegraphics[width=0.85\textwidth]{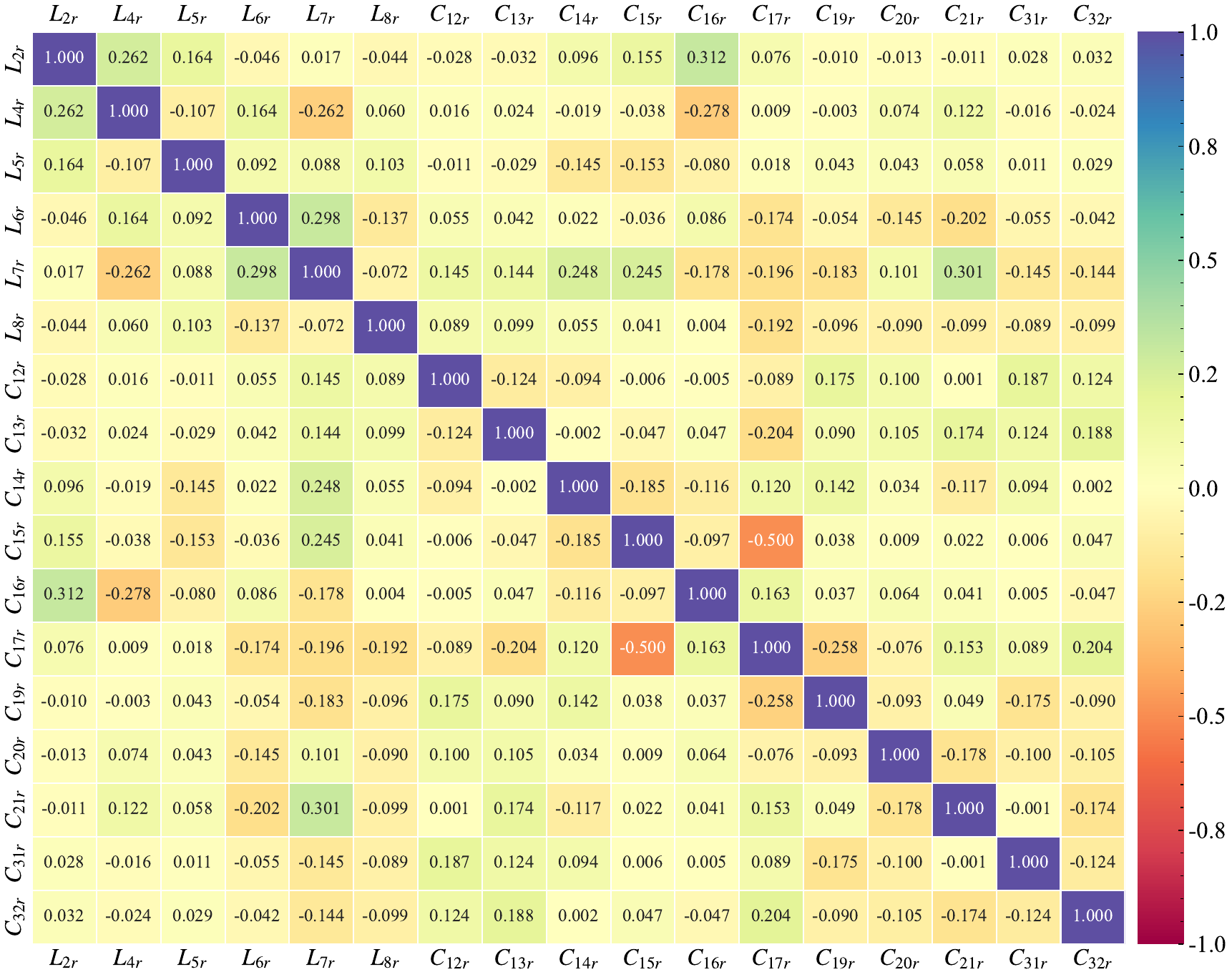}
    \caption{The correlations between the parameters obtained from the Fit 1.}
    \label{fig:corr-full-fit}
\end{figure*}
\begin{figure*}[!hpt]
    \centering
    \includegraphics[width=0.85\textwidth]{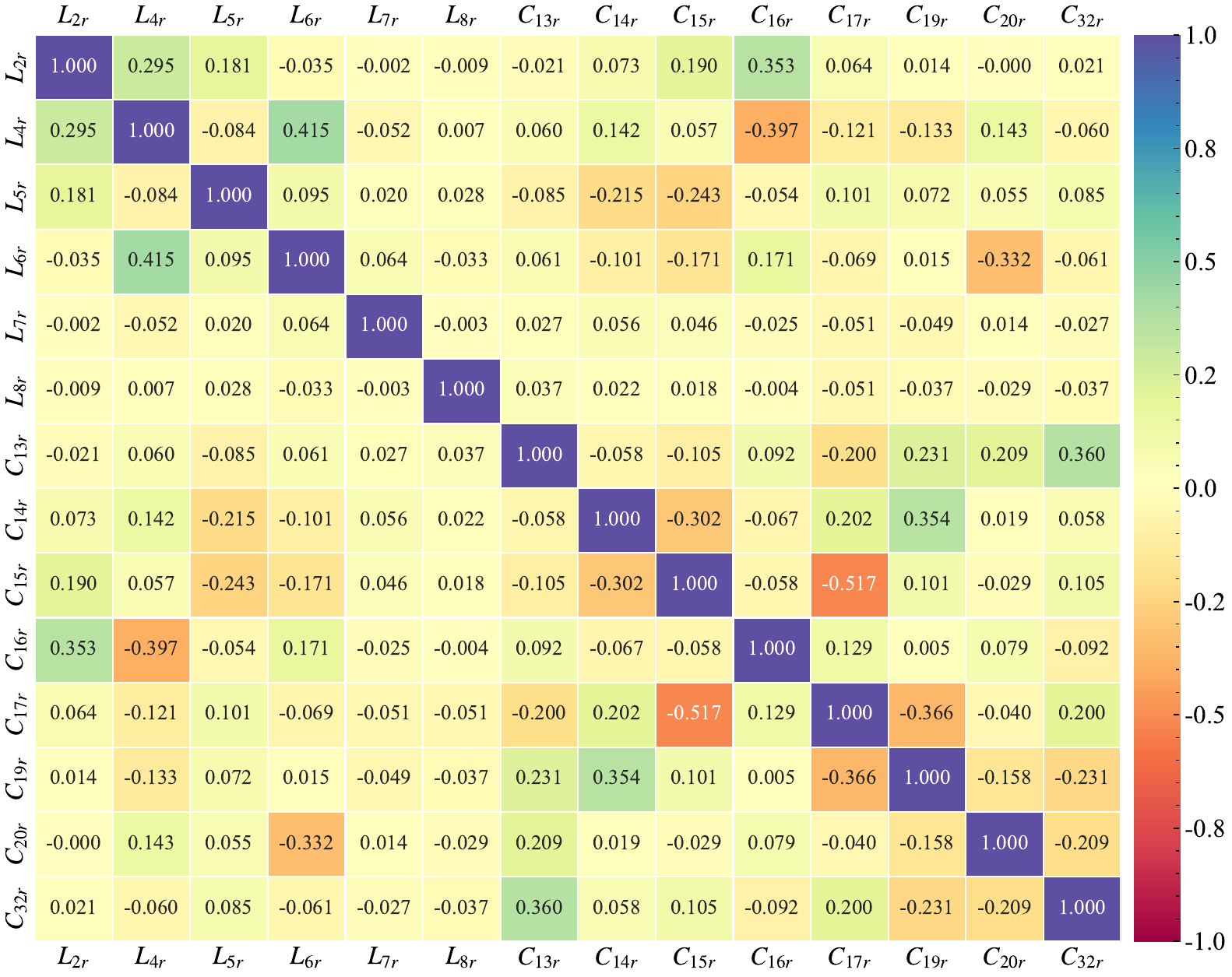}
    \caption{\label{fig:corr-fit2A}The correlations between the parameters obtained from the Fit 2A.}
    \label{fig:corr-fitA}
\end{figure*}
%\begin{figure*}[!htbp]
%    \centering
%    \caption{\label{fig:corr-Fit2B1}The correlations between the parameters obtained from the Fit 2.B1.}
 %   \includegraphics[width=0.85\textwidth]{figure/lassoB1_corr.pdf}
%\end{figure*}
\begin{figure*}[!ht]
    \centering
    \includegraphics[width=0.85\linewidth]{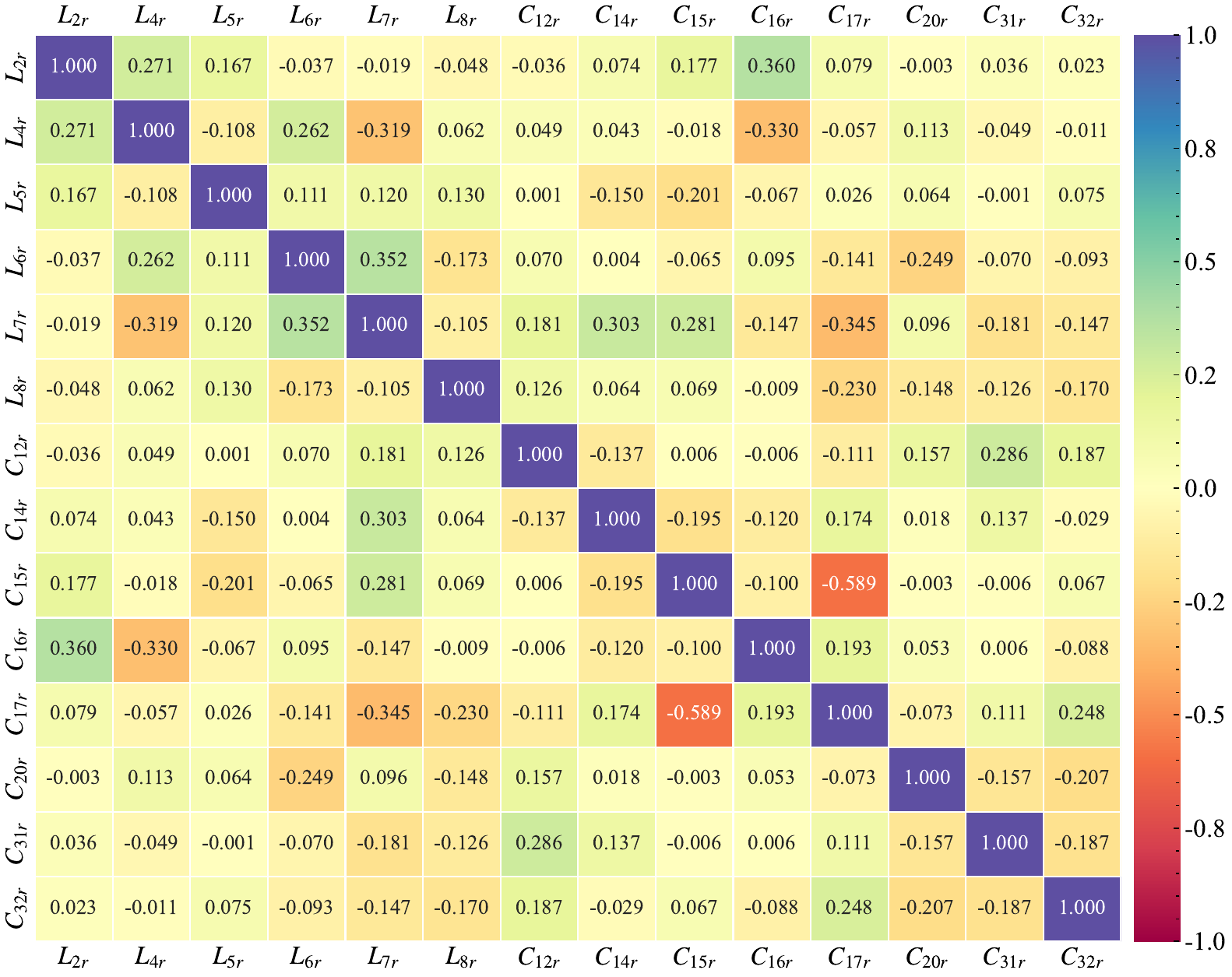}
    \caption{\label{fig:corr-fit2B}The correlations between the parameters obtained from the Fit 2B.}
\end{figure*}
%  \begin{widetext}
% % \setcounter{page}{1}
% % \setcounter{figure}{0}  
% % \setcounter{table}{0}  
% % \section{Supplemental Material}
% \appendix
% We provide the correlation matrices in the three fits.
% \section{\label{app:cor}Correlations}
% %
% \begin{figure*}[!hpt]
%     \centering
%     \includegraphics[width=0.85\textwidth]{figure/full_corr.pdf}
%     \caption{The correlations between the parameters obtained from the Fit 1.}
%     \label{fig:corr-full-fit}
% \end{figure*}
% \begin{figure*}[!hpt]
%     \centering
%     \includegraphics[width=0.85\textwidth]{figure/lassoA_corr.pdf}
%     \caption{\label{fig:corr-fit2A}The correlations between the parameters obtained from the Fit 2A.}
%     \label{fig:corr-fitA}
% \end{figure*}
% %\begin{figure*}[!htbp]
% %    \centering
% %    \caption{\label{fig:corr-Fit2B1}The correlations between the parameters obtained from the Fit 2.B1.}
%  %   \includegraphics[width=0.85\textwidth]{figure/lassoB1_corr.pdf}
% %\end{figure*}
% \begin{figure*}[!ht]
%     \centering
%     \includegraphics[width=0.85\linewidth]{figure/lassoB_corr.pdf}
%     \caption{\label{fig:corr-fit2B}The correlations between the parameters obtained from the Fit 2B.}
% \end{figure*}
%  \end{widetext}

% \bibliography{ref.bib}
%apsrev4-2.bst 2019-01-14 (MD) hand-edited version of apsrev4-1.bst
%Control: key (0)
%Control: author (8) initials jnrlst
%Control: editor formatted (1) identically to author
%Control: production of article title (-1) disabled
%Control: page (0) single
%Control: year (1) truncated
%Control: production of eprint (0) enabled
%

\end{document}